\documentclass[final,english]{bullsrsl}

\usepackage[latin1]{inputenc}
\usepackage{xcolor}
\usepackage[T1]{fontenc}

\usepackage{natbib} 

\usepackage{graphicx}

\begin{document}
\title{Asteroseismic investigation of Intermediate age Open Star Cluster NGC 2126}



\author[affil={1,2},corresponding]{Athul}{Dileep}
\author[affil={1}]{Santosh}{Joshi}
\author[affil={3}]{Patricia}{Lampens}
\author[affil={4}]{Chenglong}{Lv}
\author[affil={5}]{David}{Mkrtichian}
\author[affil={3}]{Peter}{De Cat}
\author[affil={6}]{Abhay}{Pratap Yadav}
\author[affil={1}]{Arjun}{Dawn}
\author[affil={4}]{Ali}{Esamdin}


\affiliation[1]{Aryabhatta Research Institute of Observational Sciences, Nainital 263001, Uttarakhand, India }
\affiliation[2]{Mahatma Jyotiba Phule Rohilkhand University, Bareilly, Uttar Pradesh 243006, India }
\affiliation[3]{Royal Observatory of Belgium, Ringlaan 3, B-1180 Brussels, Belgium}
\affiliation[4]{Xinjiang Astronomical Observatory, Chinese Academy of Sciences, Xinjiang 830011, People's Republic of China}
\affiliation[5]{National Astronomical Research Institute of Thailand, Chiang Mai 50180, Thailand}
\affiliation[6]{Department of Physics and Astronomy, National Institute of Technology Rourkela, Rourkela 769008, Odisha, India}


\correspondance{dileep@aries.res.in}


\maketitle

\begin{abstract}
We present a preliminary asteroseismic investigation of the pulsating stars in the intermediate-age open cluster NGC\,2126. The TESS Full Frame Images and ground-based photometry from the previous study, \citet{10.1093/mnras/staf361}, were used to identify the pulsation modes of $\delta$ Scuti member stars using period-luminosity relations and pulsation modelling. Our analysis resulted in a tentative identification up to the first radial overtone in stars N1, N2 and ZV1. Additionally, the second radial overtone was secured in N1. The disentangled pulsation and eclipse light curves of the binary star V551 Aur (V6) from our previous research were employed to study pulsation amplitude and phase modulations across the orbital period. The primary component of V6 exhibits pulsations showing amplitude and phase modulations, indicating tidal influence on the oscillations. Modelling the pulsation as the fundamental radial mode reveals that in order for the mode to be radial, the star must be significantly brighter than observed. Hence, the mode is not the fundamental radial mode or the overtones. Low-resolution spectroscopy was used to search for chemical peculiarities in the pulsating members, and we found that star ID 143 has features similar to an Am star. 
\end{abstract}





\section{Introduction}
Open clusters act as stellar laboratories where one can test theories of stellar evolution. They provide constraints on age, initial composition, metallicity and rotation rate for the member stars \citep{2024A&A...686A.142L}. The existence of pulsating stars within an open cluster makes it optimal for deriving its precise parameters via asteroseismology  \citep{Corsaro_2012}. Utilising the pulsation frequencies that are independent of distance and reddening, one can ascertain the accurate ages and other physical characteristics of the clusters. The age estimates from colour-magnitude diagrams (CMD) and asteroseismology are expected to be similar. The latter method enables us to constrain the age that is independent of the photometric errors associated with CMD \citep{2023A&A...679A..23B}.

Intermediate-age open clusters contain A- and F-type pulsators exhibiting $p$- and $g$-modes. One class of the $p$-mode pulsators, namely $\delta$ Scuti stars, exhibit a strong correlation with the star's evolutionary status through its density \citep{2014A&A...563A...7S, 2015ApJ...811L..29G}, whereas $g$-mode pulsators are important in examining the internal core rotation rate of the star \citep{10.1093/mnras/stw2717}. A fraction of A- and F-type stars also displays an unusual overabundance of heavier elements and a deficiency of lighter elements due to atomic diffusion occurring at low rotational velocities \citep{1970ApJ...160..641M}. They are known as the chemically peculiar (CP) stars \citep{1974ARA&A..12..257P}. The Am/Fm stars are CP stars characterised by weak or undetectable magnetic fields and exhibit an overabundance of Fe-peak and rare-earth elements, as well as a deficiency of light elements such as Ca and Sc. Am stars are not expected to show pulsations because atomic diffusion depletes helium in the upper atmosphere of the star. This affects the helium partial ionization zones that drive pulsations, but recent observations have reported many pulsating Am stars \citep{2011A&A...535A...3S}.

For the current study, we have performed an asteroseismic investigation of $\delta$ Scuti stars in the intermediate-age open star cluster NGC\,2126. Using the period-luminosity (P-L) relations and theoretical pulsation models, we identified the possible radial modes. Subsequently, we analyse the amplitude and phase modulations of the pulsating eclipsing binary system V551 Aur (V6), and finally, we search for pulsating CP stars in the cluster using the low-resolution spectra.

The manuscript is structured as follows: Sec.~\ref{obs} presents the observations and data analysis; Sec.~\ref{mode} details the mode identification of the member $\delta$ Scuti stars; Sec.~\ref{V6} gives an in-depth evaluation of pulsating eclipsing binary V6; Sec.~\ref{CP} provides the low-resolution spectroscopy, and finally, Sec.~\ref{con} discusses our findings and outlines our future plans for the project.

\section{Observations and Data Analysis}\label{obs}

In this study, we used the time-series data of three different sectors observed by the Transiting Exoplanet Survey Satellite (TESS) \citep{2014JAVSO..42..234R} as compiled by \citet{10.1093/mnras/staf361} for the intermediate-age open cluster NGC\,2126 utilising the TESS Full-Frame Images (FFIs). 


The low-resolution spectra for four of our pulsating member stars were observed by the Large Sky Area Multi-Object Fibre Spectroscopic Telescope (LAMOST) and downloaded from Data Release 10 (DR10) of the LAMOST archive (\url{https://www.lamost.org/dr10/v2.0/search}; \citealt{2012RAA.Cui, 2012RAA.Zhao}). 
Additionally, the low-resolution spectra for six of our pulsating members were taken with the ARIES Devasthal Faint Object Spectrograph and Camera (ADFOSC; \citealt{2009arXiv0905.0360M}).

The spectra were processed using the Image Reduction and Analysis Facility (IRAF; \url{https://iraf-community.github.io/}; \citealt{1986SPIE..627..733T}).

\section{Mode identification}\label{mode}
\subsection{Period-luminosity relations}

\begin{figure}
\centering
\includegraphics[width=0.8\textwidth]{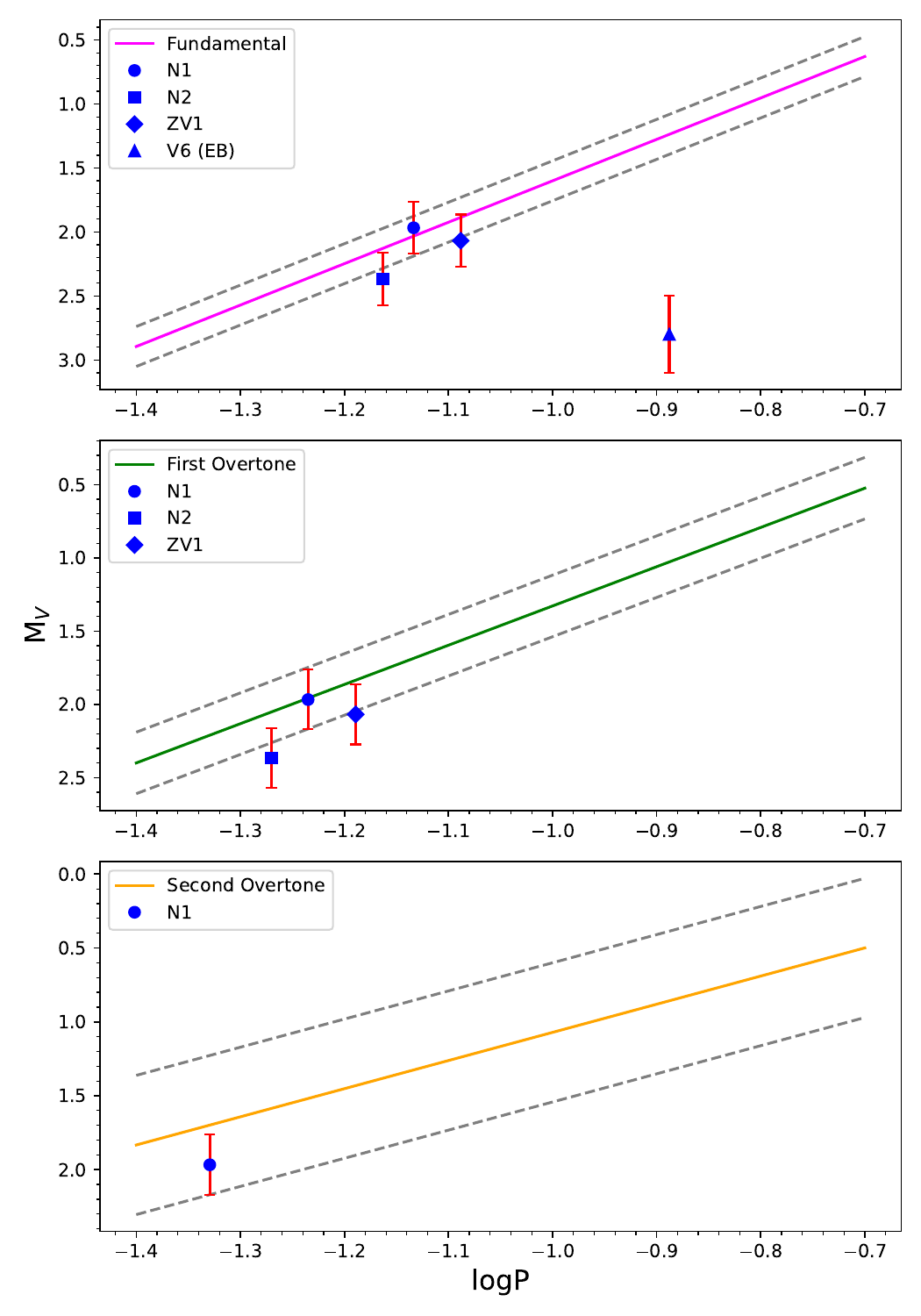}
\begin{minipage}{12cm}
\centering
\caption{P-L relations for fundamental mode (top), first overtone (middle) and second overtone (bottom) for $\delta$ Scuti stars. The x-axis denotes the logarithm of the period in days, while the y-axis indicates the absolute magnitude ($M_{V}$) of the star. The markers represent the frequencies for different stars, as shown in the legend. The dashed lines represent the minimum 3 $\sigma$ uncertainties in the relation.}
\label{PL}
\end{minipage}
\end{figure}

$\delta$ Scuti stars follow a P-L relation by which we can estimate the excited radial modes ($\ell = 0$) of the star given its intrinsic brightness. The radial modes of low order are most likely to be excited to a higher amplitude in $\delta$ Scuti stars with slow to moderate rotation, particularly observed in the case of High Amplitude Delta Scuti stars (HADS) \citep{2000ASPC..210..373M}. We considered the frequencies exhibiting the largest amplitude across all TESS sectors to be radial modes and verified if these frequencies adhered to the P-L relation \citep{Poro_2024} for the fundamental radial mode or the overtones (see Fig.~\ref{PL}). The absolute magnitudes ($M_{V}$) of the stars were estimated from Gaia parallaxes and cluster extinction, which were subsequently transformed into bolometric magnitudes using the empirical relations given by \citet{2010torres} to obtain their luminosities. The stars N1, N2, and ZV1 conformed to the P-L relation up to the first overtone, while N1 also satisfied the second overtone. The Eclipsing Binary (EB) V6, however, was shown to deviate significantly from the P-L relation, even for the radial overtones. 
That is, the single dominant frequency observed in V6 lies outside the domain of radial modes for its brightness. The basic physical parameters and identified radial modes of the stars are tabulated in Table \ref{Tab:Table 2}. They are used to produce the best theoretical models for the studied stars, as described in the next subsection.

\begin{table}
\centering
\begin{minipage}{160mm}  
\centering
\caption{The basic parameters, effective temperature ($T_{\rm eff}$), surface gravity ($\log g$) and luminosity ($\log L$) of the $\delta$ Scuti stars, along with the identified radial fundamental, first overtone and second overtone mode frequencies. The parameters of V6 are marked with `*' since they are derived from the composite medium-resolution spectra and the observed integrated magnitude of the EB star. The $T_{\rm eff}$ of other stars was calculated from ground-based multiband photometry.}
\label{Tab:Table 2}
\end{minipage}
\begin{tabular}{ccccc}
\hline
 \textbf{IDs} &\textbf{N1} & \textbf{N2} & \textbf{ZV1}& \textbf{V6} \\
\hline
R.A. (deg) &90.661139  &90.614056   & 90.637472 & 90.658250 \\
Dec. (deg) &49.879417 & 49.845361 & 49.880111&49.884250 \\
\hline
       $T_{\rm eff}$ (K) & 7130(260)  & 7140(270)&7080(250)& 6460(360)$^{*}$\\
        $\log g$ & - &- &-&3(1)$^{*}$\\
        $\log L$ &1.08(6)&0.92(6)&1.06(6)&0.8(1)$^{*}$\\
        F0 (d$^{-1}$)&13.5969(9) &14.5554(7) &12.2504(6)&-\\
        F1 (d$^{-1}$)& 17.177(2)&18.627(2)&15.458(1)&-\\
        F2 (d$^{-1}$)& 21.203(1) &-&-&-\\
        F0/F1 ($\sim$0.76-0.79)$^{+}$& 0.79 &0.78 &0.79 &-\\
        F0/F2 ($\sim$0.61-0.63)$^{+}$& 0.64 &- &- &-\\
        F1/F2 ($\sim$0.8)$^{\times}$& 0.81&- &- &-\\
        
\hline
\multicolumn{5}{l}{\small $^{+}$\citet{1979ApJ...227..935S}, $^{\times}$\citet{2021AcA....71..189S}}
\end{tabular}
\end{table}

\subsection{MESA and GYRE}
\begin{figure}
\centering
\includegraphics[width=0.8\textwidth]{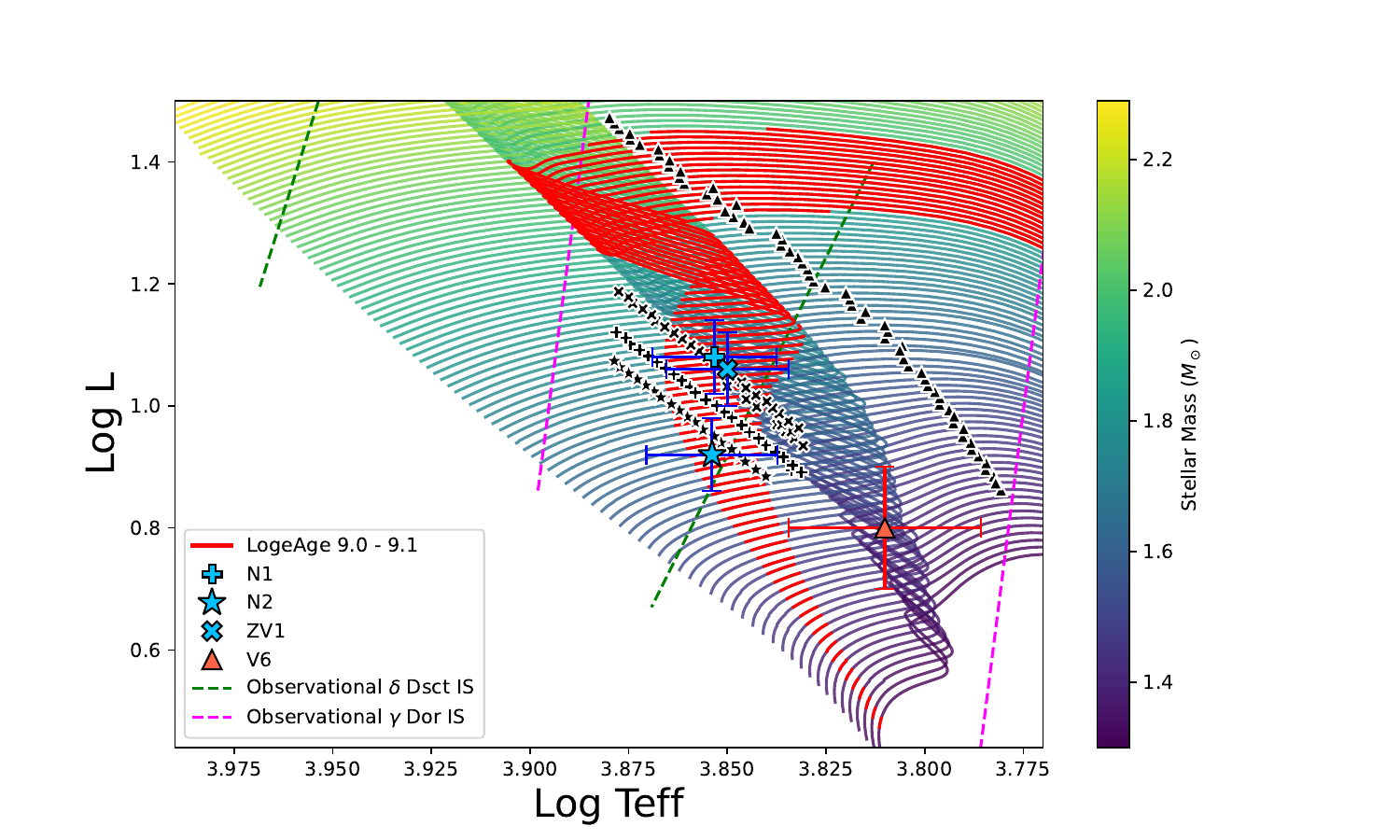}
\begin{minipage}{12cm}
\centering
\caption{MESA evolution tracks for varying masses are depicted using colour code. The member $\delta$ Scuti stars, based on their $T_{\rm eff}$ and $\log L$ obtained from their B-V colour and Gaia parallaxes, are marked. The red regions mark the $\log t$ in the range from 9.0 to 9.1 of the cluster. The observational $\delta$ Scuti \citep{murphyis2019} and $\gamma$ Doradus \citep{2002MNRAS.333..251H} instability strips are depicted with dashed lines. The black markers are the best models for each star.}
\label{HRD}
\end{minipage}
\end{figure}

We constructed a grid of stellar models using the Modules for Experiments in Stellar Astrophysics (\texttt{MESA}) stellar evolution code (version r22.05.1; \citealt{Paxton2011, Paxton2013, Paxton2015, Paxton2018, Paxton2019, Jermyn2023}).The metallicity of the cluster was set to Z=0.016, the value taken from \citet{10.1093/mnras/staf361}. Models with a mass range from 1.3 to 2.3 M$_{\odot}$ in increments of 0.01 M$_{\odot}$ were constructed. The associated radial mode frequencies were computed using the GYRE stellar oscillation code \citep{2013MNRAS.435.3406T, 2018MNRAS.475..879T}. The modes identified from the preceding subsection were used to find the best models. The best-fit models were identified by minimizing the $\chi^2$ between the observed and model frequencies. For each star, corresponding to the identified modes, the best models are illustrated in Fig.~\ref{HRD}. The red regions in Fig.~\ref{HRD} indicate the evolutionary ages $\log t(year) = 9.0 - 9.1$ for the open cluster. The age of the cluster can be constrained with higher accuracy once the best model for each star is identified. The inlists used for creating the grid are available at Zenodo, at this \href{https://zenodo.org/records/20843006}{\textcolor{blue}{link}}.

The stars N2 and ZV1 appear to align closely with the best-fit models. However, star N1 appears to possess a slightly larger luminosity. A more meticulous analysis is required to determine the optimal models for each star. In the case of V6, the best-fit models were selected assuming that the (single) dominant mode is the radial fundamental mode. In such a scenario, the brightness appears to be underestimated by approximately 0.4 dex. However, the cluster extinction calculated from CMD (E$_{B-V}$=0.2), or the extinction calculated from the interstellar Na D lines (E$_{B-V}$=0.29) from medium-resolution spectra of the star, does not account for the diminished luminosity. The position of V6 in the HR diagram places it within the observational $\gamma$ Doradus instability strip, but the pulsation frequency of 7.7266 d$^{-1}$ is highly unlikely for a g-mode unless the star is a very fast rotator. The medium-resolution spectroscopy by \citet{10.1093/mnras/staf361} reported $\upsilon\sin i$ = 44 km\,s$^{-1}$, which means the star is likely a moderate rotator.  

In order to study the effects of orbital motion on the stellar pulsations, we used the TESS light curves to investigate the amplitude and phase modulation of pulsations over the orbital phase, and the results are presented in the next section.

\section{Amplitude and phase modulations in V551 Aur (V6)}\label{V6}

The target V551 Aur is an eclipsing binary member of NGC\,2126 with a pulsating component. \citet{2018chehlah} performed a detailed study of the  open cluster NGC\,2126 and particularly V551 Aur using the ground-based photometry. Recently, \citet{10.1093/mnras/staf361} utilised the TESS Full Frame Images (FFIs) light curves and disentangled eclipse and pulsation features by conducting simultaneous modelling of the eclipses, considering pulsations as a Gaussian process. 

\begin{figure}
\centering
\includegraphics[width=0.8\textwidth]{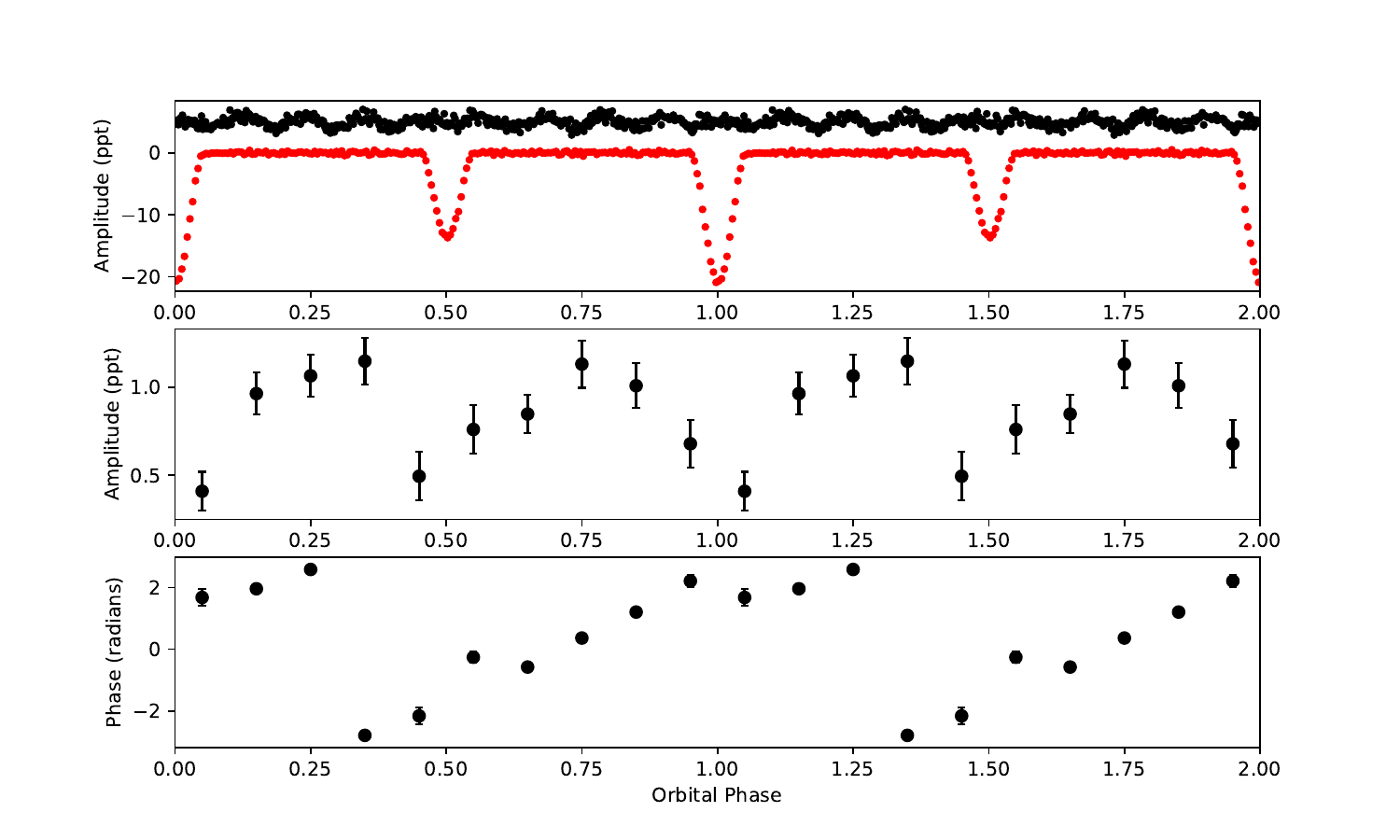}
\begin{minipage}{12cm}
\centering
\caption{\textit{Top Panel:} Disentangled light curves of the eclipses (red) and pulsation (black), phase folded with the orbital period and binned to 400 bins. \textit{Middle Panel:} The variation of amplitude of the pulsation with orbital phase. \textit{Bottom Panel:} Variation of phase of the pulsation with orbital phase.}
\label{modulation}
\end{minipage}
\end{figure}

The disentangled pulsation light curves were phase-folded using the orbital period of the system and binned into 400 phase bins. Subsequently, we fitted a sinusoidal function to each phase segment (width = 0.1) with the pulsation frequency 7.7266 d$^{-1}$ and determined the amplitude and phase. The reference point for phase variations was chosen as the epoch of primary minima ($t_0$=2452307.410 BJD). The amplitude and phase light curves are depicted in Fig.~\ref{modulation}, which exhibits clear periodic variations. The pulsation amplitude shows minima during both primary and secondary eclipses, while the pulsation phase passes through zero near the quadratures (phases 0.25 and 0.75). The phase flip at 0.25 is abrupt, whereas that at 0.75 is gradual. This phase flipping usually implies that the star is tidally distorted by the companion \citep{2025ApJ...979...80F}. A more detailed study is necessary to understand the geometry of the system and the nature of the pulsations. Note that these pulsations closely align with the $9^{th}$ harmonic of the orbital frequency \citep[e.g.,][]{2018chehlah}.

\section{Chemical Peculiarity}\label{CP}

Normal late A- to F-type stars display prominent Ca {\sc ii K} (3934 \AA) lines. However, in peculiar Am stars, the Ca {\sc ii K} lines are unusually weak for their spectral class, which facilitates the identification of CP1-type peculiarities even in low-resolution spectra. In addition, a weak Ca {\sc i} (4226 \AA) line and the weakening of lines in the range 4395-4400 \AA, due to significant contribution from Sc {\sc ii}, are other prominent features visible in low-resolution spectra \citep{hummerich2021lamost}.

\begin{figure}
\centering
\includegraphics[width=0.8\textwidth]{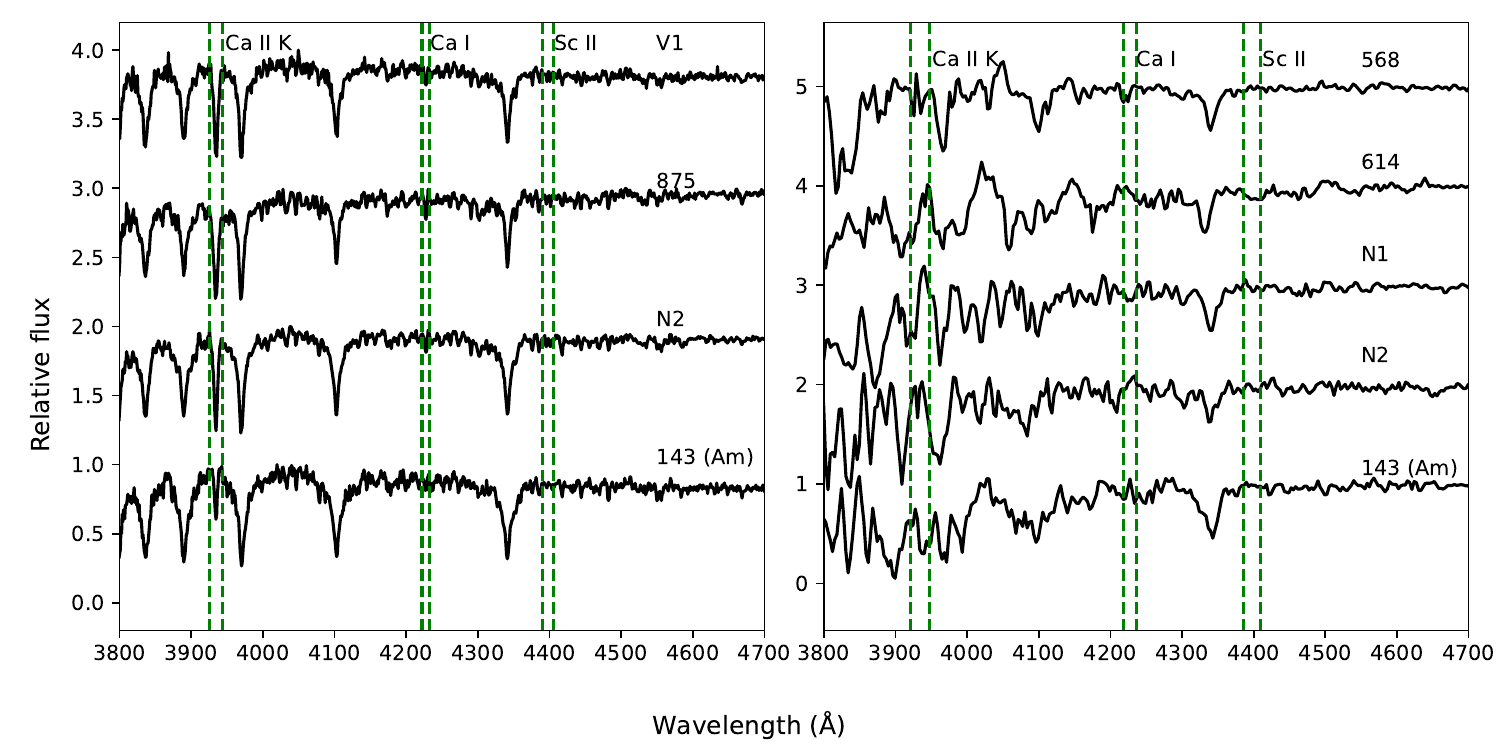}
\begin{minipage}{12cm}
\centering
\caption{LAMOST (\textit{left}) and ADFOSC (\textit{right}) low-resolution spectra of pulsating stars in NGC\,2126. The spectra of each target are offset by a factor of 1 for visibility. The dashed lines represent the regions where Ca {\sc ii K}, Ca {\sc i}, and Sc {\sc ii} lines are present.}
\label{LRS_spec}
\end{minipage}
\end{figure}

In the LAMOST spectra of pulsating stars illustrated in Fig.~\ref{LRS_spec} (left), the star 143 exhibits noticeably weak Ca {\sc ii K} and Ca {\sc i} lines and shows line weakening at the Sc {\sc ii} region and is likely to be an Am star, while all the other stars have pronounced Ca absorption lines. Stars 143 and N2 were additionally studied with ADFOSC. The spectra (Fig.~\ref{LRS_spec} (right)) show the line weakening in the Sc {\sc ii} region, though the Ca {\sc ii K} and Ca {\sc i} regions are unreliable due to inadequate signal-to-noise ratio (SNR). In the near future, the ADFOSC spectra will be used to confirm the spectral type of the stars.

\section{Discussions and Future plan}\label{con}

The P-L relations were used for initial mode identification of the target stars, followed by MESA and GYRE to find the best models. We tentatively identified the radial modes up to the first overtones in N1, N2, and ZV1, and the second overtone in N1. Further, we will explore the effect of metallicity and initial rotation rates on the models to get more precise models to constrain the cluster age.

The amplitude and phase modulations of V6 suggest tidal interactions. P-L relations and pulsation modelling suggest that the single dominant mode excited in V6 is not the radial fundamental mode or the overtones. To understand the real nature of this pulsation, we require precise orbital parameters of the binary system, which can only be determined by integrating spectroscopic observations with the photometric eclipses.

The low-resolution spectra were used for searching peculiarities. The LAMOST spectra proved to be more reliable, given better SNR over Ca {\sc ii K}, Ca {\sc i} and Sc {\sc ii} regions. The ADFOSC spectra have good SNR only in the Sc {\sc ii} region. On the basis of the analysis of these spectra, we classified star 143 as a probable Am candidate. 

The current study presents our initial results of the asteroseismic investigation of the open cluster NGC\,2126. In the near future, we will extend this exploration to study the pulsating stars in detail to constrain the age of the cluster.

\begin{acknowledgments}
This work is a result of the Belgo-Indian Network for Astronomy and Astrophysics (BINA), which promotes astronomical collaborations between Indian and Belgian partners and received financial support from the International Division of the Department of Science and Technology (DST, Government of India) and the Belgian Federal Science Policy Office (BELSPO, Government of Belgium) during the period 2016-2023 through different bilateral projects. AD acknowledges the financial assistance obtained from the DST-INSPIRE Fellowship Programme (DST/INSPIRE Fellowship/2020/IF200245). SJ acknowledges the BRICS grant for the project DST/IC/BRICS/2023/5. This work incorporates data acquired from the TESS mission, sourced from the Barbara A. Mikulski Archive for Space Telescopes (MAST) at the Space Telescope Science Institute (STScI) and the SIMBAD databases managed by the NASA Explorer Program and CDS in Strasbourg, France, respectively. Guoshoujing Telescope (the Large Sky Area Multi-Object Fiber Spectroscopic Telescope LAMOST) is a National Major Scientific Project built by the Chinese Academy of Sciences. Funding for the project has been provided by the National Development and Reform Commission. LAMOST is operated and managed by the National Astronomical Observatories, Chinese Academy of Sciences. The MESA EOS is a blend of the OPAL \citep{Rogers2002}, SCVH \citep{Saumon1995}, FreeEOS \citep{Irwin2004}, HELM \citep{Timmes2000}, PC \citep{Potekhin2010}, and Skye \citep{Jermyn2021} EOSes. Radiative opacities are primarily from OPAL \citep{Iglesias1993, Iglesias1996}, with low-temperature data from \citet{Ferguson2005} and the high-temperature, Compton-scattering dominated regime by \citet{Poutanen2017}.  Electron conduction opacities are from \citet{Cassisi2007} and \citet{Blouin2020}. Nuclear reaction rates are from JINA REACLIB \citep{Cyburt2010}, NACRE \citep{Angulo1999} and additional tabulated weak reaction rates \citet{Fuller1985, Oda1994, Langanke2000}.  Screening is included via the prescription of \citet{Chugunov2007}. Thermal neutrino loss rates are from \citet{Itoh1996}.
\end{acknowledgments}


\begin{furtherinformation}

\begin{orcids}

  \orcid{0000-0002-2011-6233}{Athul}{Dileep}
  \orcid{0009-0007-1545-854X}{Santosh}{Joshi}
  \orcid{0000-0002-7034-4912}{Patricia}{Lampens}
  \orcid{0000-0001-6354-1646}{Chenglong}{Lv}
  \orcid{0000-0001-5094-3910}{David}{Mkrtichian}
  \orcid{0000-0001-5419-2042}{Peter}{De Cat}
  \orcid{0000-0001-8262-2513}{Abhay}{Pratap Yadav}
  \orcid{0009-0005-3400-6915}{Arjun}{Dawn}
  \orcid{0000-0003-1845-4900}{Ali}{Esamdin}

\end{orcids}

\begin{authorcontributions}
Collective efforts were made by all the co-authors with relevant contributions.
\end{authorcontributions}

\begin{conflictsofinterest}
The authors declare that there is no conflict of interest.
\end{conflictsofinterest}

\end{furtherinformation}



%

\bibliographystyle{bullsrsl-en}

\bibliography{extra.bib}

\end{document}